\documentclass[12pt,preprint]{aastex}

\shorttitle{Galactic Wind in NGC 253 with the Kyoto3DII}
\shortauthors{Matsubayashi et al.}

\begin{document}

\title{Galactic Wind in the Nearby Starburst Galaxy NGC 253 Observed
with the Kyoto3DII Fabry-Perot Mode}

\author{K. Matsubayashi\altaffilmark{1}, H. Sugai\altaffilmark{1},
T. Hattori\altaffilmark{2}, A. Kawai\altaffilmark{1}, S. Ozaki\altaffilmark{3},
G. Kosugi\altaffilmark{4}, T. Ishigaki\altaffilmark{5},
A. Shimono\altaffilmark{1}}
\altaffiltext{1}{Department of Astronomy, Kyoto University, Sakyo-ku, Kyoto
606-8502, Japan; kazuya@kusastro.kyoto-u.ac.jp}
\altaffiltext{2}{Subaru Telescope, National Astronomical Observatory of Japan,
650 North A'ohoku Place, Hilo, Hawaii 96720, USA}
\altaffiltext{3}{Okayama Astrophysical Observatory, National Astronomical
Observatory of Japan, Kamogata, Okayama 719-0232, Japan}
\altaffiltext{4}{National Astronomical Observatory of Japan, Mitaka, Tokyo
181-8588, Japan}
\altaffiltext{5}{Asahikawa National College of Technology, Asahikawa, Hokkaido
071-8142, Japan}
\begin{abstract}

We have observed the central region of the nearby starburst galaxy NGC 253
with the Kyoto Tridimensional Spectrograph II (Kyoto3DII) Fabry-Perot mode
in order to investigate the properties of its galactic wind.
Since this galaxy has a large inclination, it is easy to observe its galactic
wind.
We produced the H$\alpha$, [\ion{N}{2}]$\lambda$6583, and
[\ion{S}{2}]$\lambda\lambda$6716,6731 images, as well as those line ratio maps.
The [\ion{N}{2}]/H$\alpha$ ratio in the galactic wind region is larger than
those in \ion{H}{2} regions in the galactic disk.
The [\ion{N}{2}]/H$\alpha$ ratio in the southeastern filament, a part of the
galactic wind, is the largest and reaches about 1.5.
These large [\ion{N}{2}]/H$\alpha$ ratios are explained by shock
ionization/excitation.
Using the [\ion{S}{2}]/H$\alpha$ ratio map, we spatially separate the galactic
wind region from the starburst region.
The kinetic energy of the galactic wind can be sufficiently supplied by
supernovae in a starburst region in the galactic center.
The shape of the galactic wind and the line ratio maps are non-axisymmetric
about the galactic minor axis, which is also seen in M82.
In the [\ion{N}{2}]$\lambda$6583/[\ion{S}{2}]$\lambda\lambda$6716,6731 map,
the positions with large ratios coincide with the positions of star clusters
found in the {\it Hubble Space Telescope} ({\it HST}) observation.
This means that intense star formation causes strong nitrogen enrichment
in these regions.
Our unique data of the line ratio maps including [\ion{S}{2}] lines have
demonstrated their effectiveness for clearly distinguishing between shocked gas
regions and starburst regions, determining the extent of galactic wind and its
mass and kinetic energy, and discovering regions with enhanced nitrogen
abundance.

\end{abstract}


\section{INTRODUCTION}

Galactic scale outflows (galactic winds) from galactic disks are common
phenomena in the universe: they are found both in nearby galaxies (e.g.,
\citealt{Strickland:2004,Tullmann:2006}) and in high-z galaxies
(e.g., \citealt{Shapley:2003,Tapken:2007}).
Their main sources of energy are stellar winds, supernovae, or active galactic
nuclei (AGNs).
Their asymmetric and complex structures (e.g., M82; \citealt{Shopbell:1998},
NGC 253; \citealt{Sugai:2003}) are investigated by numerical simulations
(e.g., \citealt{Cooper:2008}).
Galactic winds are considered to impact on galaxies and their environments.
\citet{Kobayashi:2007} takes galaxy-scale outflows into their semi-analytical
model of galaxy formation as a parameter, which reproduces the observed
Ly$\alpha$ luminosity functions of Ly$\alpha$ emitters in high-z.
Recently, \citet{Sato:2008} has suggested from their multi-wavelength
observation data that galactic winds may carry away interstellar gas in host
galaxies and may quench star formation.
Galactic winds from dwarf galaxies with gas masses of a few $\times 10^8
M_{\odot}$ are suggested to be the major pollutants of the intergalactic
medium (IGM) through an analytical/numerical model \citep{Ferrara:2000}.
\citet{Veilleux:2005} and \citet{Bland:2007}, in reviews of the galactic
winds, summarize the properties and significance of galactic winds.

Despite the importance of galactic winds, there are few galaxies where the
basic physical values (e.g., mass and kinetic energy) of the galactic winds
were derived from spatially resolved data.
It is important to analyse spatially resolved data because galactic winds
have extremely complex structures: e.g., non-axisymmetry about the minor axis
of a galactic disk is found in M82 \citep{Shopbell:1998} and NGC 253
\citep{Sugai:2003}.
The Fabry-Perot Interferometer, which enables us to subtract the continuum
precisely and to produce pure line images, is a powerful method to observe
galactic winds.
The galactic winds in M82 \citep{Shopbell:1998} and NGC 1482 \citep
{Veilleux:2002} were observed with Fabry-Perot instruments, and their masses,
kinetic energies, and excitation sources of them were estimated from
the [\ion{N}{2}]$\lambda$6583/H$\alpha$ line ratio maps.
In spite of the fact that this method is especially effective, there were
only three examples where the properties of galactic winds in edge-on
galaxies were discussed in detail from line ratio maps.

NGC 253 is one of the most famous galaxies because it is a nearby \citep
[D=3.5 Mpc:][]{Rekola:2005} edge-on \citep [inclination=78$^{\circ}$:][]
{Pence:1980} starburst galaxy.
Details of this galaxy have been investigated in many wavelengths such as
X-ray \citep{Strickland:2000}, optical \citep{Schulz:1992, Watson:1996},
infrared \citep{Sams:1994,Sugai:2003}, sub-mm \citep{Sakamoto:2006}, and
radio \citep{Turner:1985, Ulvestad:1997}.
In NGC 253, the galactic wind was found in H$\alpha$ \citep[e.g.,][]
{McCarthy:1987,Schulz:1992} and in X-ray \citep[e.g.,][]{Strickland:2000},
and the properties of the galactic wind were discussed in them. 
The expansion velocity is estimated through slit spectroscopic observation
in optical \citep{Schulz:1992}.
\citet{Strickland:2000} discuss the spatial structure of the outflow from
the azimuthal surface brightness profiles of the X-ray and the H$\alpha$.

We have observed the central region of NGC 253 with the Fabry-Perot(FP)
mode of the Kyoto Tridimensional Spectrograph II \citep[Kyoto3DII:][]
{Sugai:2000, Sugai:2002, Sugai:2004a} mounted on the Subaru Telescope.
The Kyoto3DII is a unique instrument that enables us to carry out the optical
Fabry-Perot observations on an 8-meter class telescope and to observe weak
emission lines.
The observations were carried out during the test observations of the
Kyoto3DII, and this is the first result of the Kyoto3DII FP mode.
%
%
From these data, we have obtained the [\ion{N}{2}]$\lambda$6583/H$\alpha$,
[\ion{S}{2}]$\lambda\lambda$6716,6731/H$\alpha$, and [\ion{N}{2}]/[\ion{S}{2}]
line ratio maps.
This unique set of line ratio maps allows us to clearly distinguish between
shocked gas regions and starburst regions, to determine the extent of galactic
wind as well as its mass and kinetic energy, and to discover regions with
enhanced nitrogen abundance.

We describe our observations and data reduction in Section \ref{sec:obs}.
We show the H$\alpha$ and continuum images and the line ratio maps, and
compare our images with the X-ray and the radio images in Section
\ref{sec:results}.
We discuss the properties of the galactic wind and stellar clusters from the
continuum and H$\alpha$ images and the line ratio maps in Section
\ref{sec:discuss}.
We summarize our conclusions in Section \ref{sec:summary}.

\section{OBSERVATIONS AND DATA REDUCTION}
\label{sec:obs}

We have observed the central region of NGC 253 with the Fabry-Perot(FP)
mode of the Kyoto3DII \citep{Sugai:2000, Sugai:2002, Sugai:2004a} on 28
August 2002 during the test observation of the Kyoto3DII.
It was mounted on the Cassegrain focus of the Subaru Telescope at Mauna Kea.
This mode uses an ET--50 etalon made by Queensgate Instruments.
The field of view is $\sim$ 1\arcmin.9 $\times$ 1\arcmin.9, with the spatial
sampling of 0\arcsec.112 after 2 $\times$ 2 on-chip binning.
The spatial resolution was 1\arcsec.0 and the spectral resolution was 21 \AA.
For each of the H$\alpha$+[\ion{N}{2}]$\lambda\lambda$6548,6583 and 
[\ion{S}{2}]$\lambda\lambda$6716,6731 sets, we obtained 5 on-band frames and 1
off-band frame.
The central wavelengths of the transmission curve of etalon for the
H$\alpha$+[\ion{N}{2}] set were 6556 \AA, 6568 \AA, 6580 \AA, 6592 \AA, 6604
\AA (on-band), and 6502 \AA (off-band), while those of the
[\ion{S}{2}]$\lambda\lambda$6716,6731 set were 6717 \AA, 6729 \AA, 6741 \AA,
6754 \AA, 6766 \AA (on-band), and 6650 \AA (off-band).
The exposure time of each frame was 240 seconds.
The bias subtraction and the flat fielding have been performed for the target
object frames and the standard star frames, and then the flux calibration
has been done.
For reducing the observing time, flat and standard star frames were not
obtained at all the observed wavelengths.
Therefore, we have used the flat frames of the nearest wavelength
for the flat fielding and the interpolated data for the flux calibration.
The flat flames in 6502 \AA, 6568 \AA, and 6650 \AA~were similar to each
other: $\sim$ 20 \% brighter at the frame center compared with at the edge.
The flat flames in longer wavelengths, including in 6729 \AA, 6742 \AA,
6749 \AA, and 6757 \AA, had a different flat pattern: darker at the frame
center.
In the longest wavelength, it was $\sim$ 30 \% darker at the center.
The resultant uncertainty caused by the flat fielding, including the
interpolation, is negligible in shorter wavelengths, and within a few percent
in longer wavelengths.
We have removed cosmic rays, matched the spatial resolution to the worst
frame, and subtracted the off-band frames from the on-band frames.
Eventually, the frames including only emission lines were obtained.
The sky subtraction has not been performed in each frame because objects are
observed in the whole field of view.
This affects the continuum image, but does not affect the line images because
the sky emission has been subtracted through the process of the off-band
subtraction.

\begin{figure}
\epsscale{1.00}
\plottwo{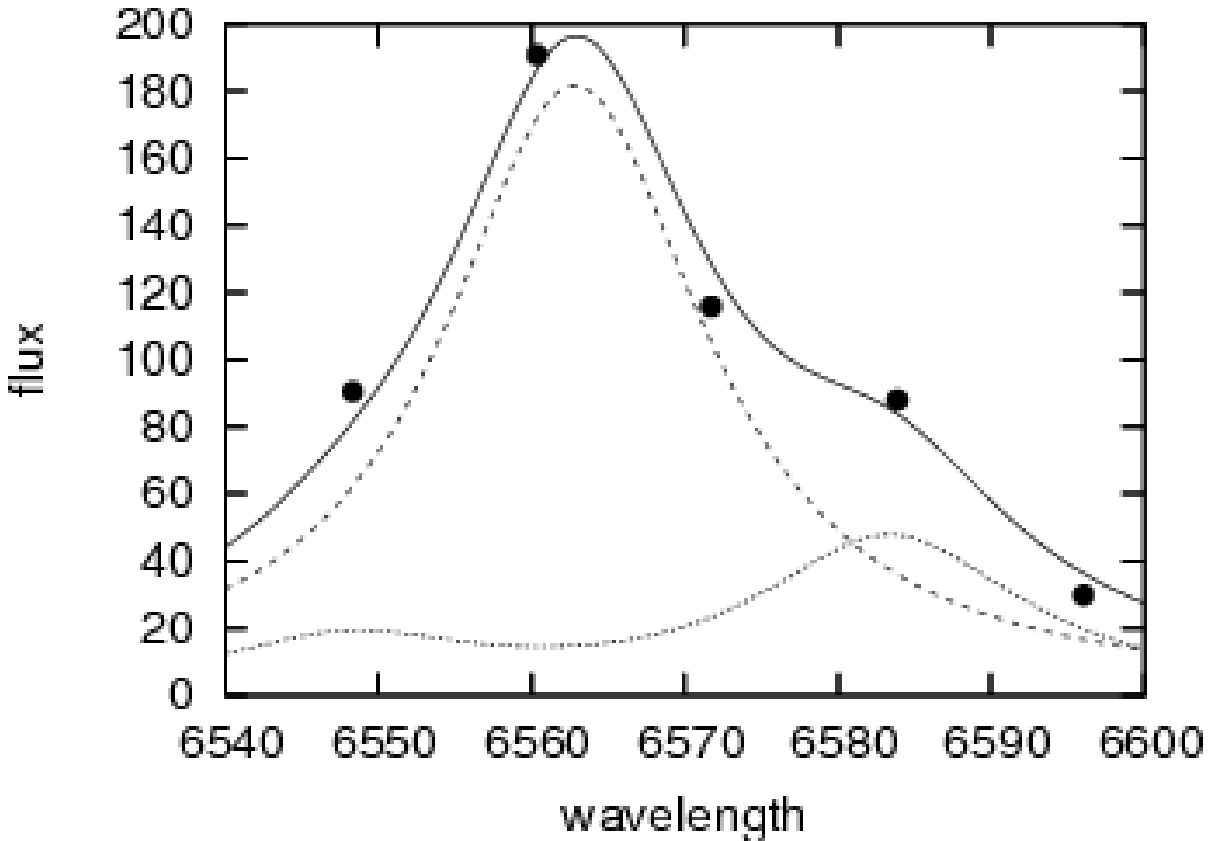}{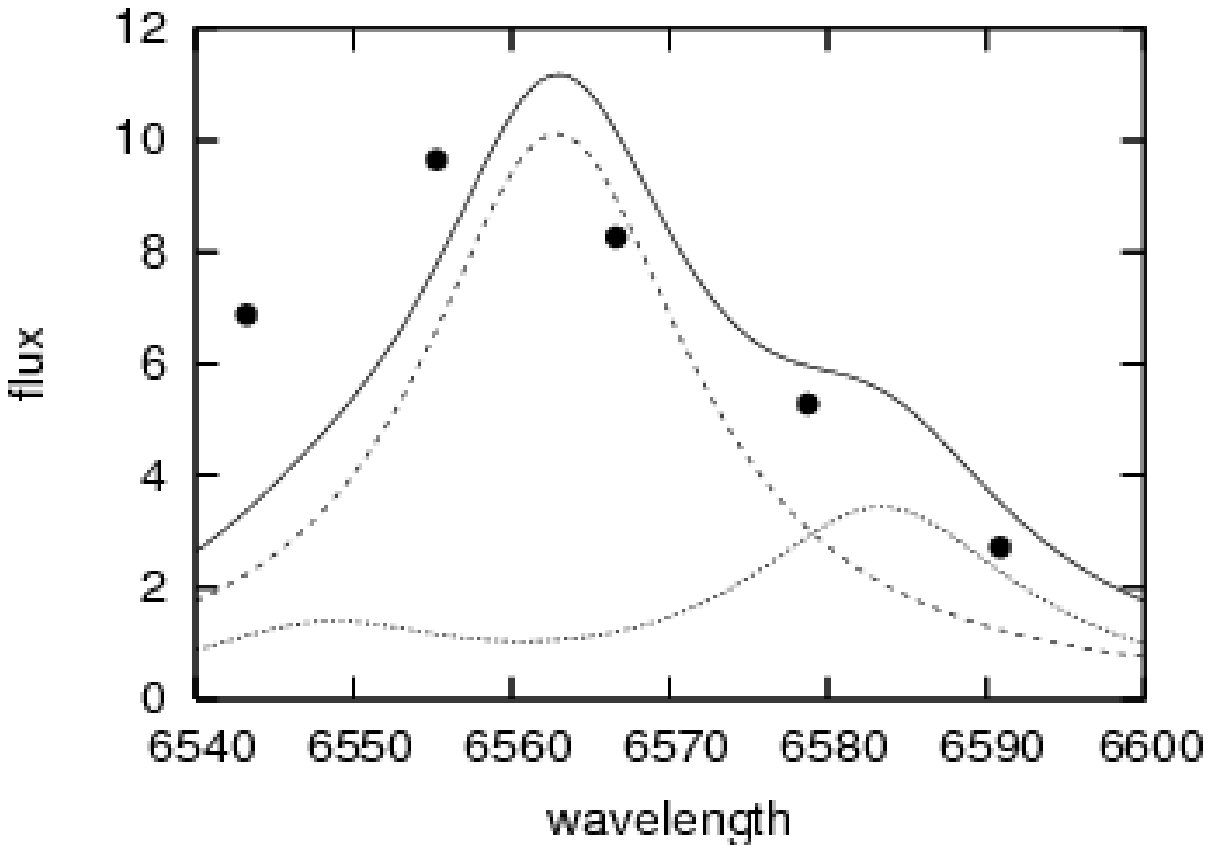}
\caption{Good ($\it{left~panel}$) and poor examples ($\it{right~panel}$) of
fitting of the H$\alpha$+[\ion{N}{2}] set.
The flux is shown as a function of wavelength in units of \AA.
The observed values are plotted in filled circles.
The dashed line and dotted line represent the fluxes from the H$\alpha$ and
the [\ion{N}{2}]$\lambda\lambda$6548,6583, respectively, as a function of the
central wavelength of the transmission curve of etalon.
The solid line is sum of the dashed line and dotted line.
\label{fig:n2-fit-example}}
\end{figure}

Because of the low spectral resolution, the H$\alpha$ and the
[\ion{N}{2}]$\lambda\lambda$6548,6583 in the H$\alpha$+[\ion{N}{2}] set, as
well as the [\ion{S}{2}]$\lambda\lambda$6716,6731 in the [\ion{S}{2}] set,
were blended.
In order to decompose them, we have fitted these lines pixel by pixel with
Airy functions, which are the transmission curve of the Fabry-Perot
Interferometer \citep{Bland:1989}.
Figure \ref{fig:n2-fit-example} shows examples of the fitting.
The residual ratios were calculated by dividing the total residuals by the
total flux of the H$\alpha$ plus [NII]$\lambda\lambda$6548,6583, and they were
5\% and 30\% in good and poor examples, respectively.
In poor fitting, line decomposing may be inaccurate, and the flux
ratio may have a large error. 
However, the fraction of poor examples was smaller than 1/100, and
much smaller in the regions we discuss below, and we neglected them.
The average residual ratio for a single pixel in the discussed regions was
15\%.
As free parameters, we took the fluxes of the H$\alpha$ and 
[\ion{N}{2}]$\lambda$6583 in the H$\alpha$+[\ion{N}{2}] set, and those of the
[\ion{S}{2}]$\lambda$6716 and [\ion{S}{2}]$\lambda$6731 in the [\ion{S}{2}]
set.
We assumed that the flux of the [\ion{N}{2}]$\lambda$6548 was one third of
that of the [\ion{N}{2}]$\lambda$6583.
For the fitting, we fixed the velocity center so that it corresponded to
the heliocentric velocity of NGC 253 of 251 km s$^{-1}$ \citep{Watson:1996}.
We also fixed the intrinsic velocity dispersion as zero because it was
much smaller than the instrumental profile.
The main source of the uncertainties for the line ratio maps is not
flux calibration but line fitting, because the sky transmittance was
stable during the observation for NGC 253.
In order to estimate the uncertainties of the fitting, we considered
pseudo-observation data: they were obtained with the same "observing"
parameters as actual observations from the spectra which include maximum
velocity variation, velocity dispersion, and line-splitting observed in
slit-spectroscopy \citep{Schulz:1992}.
The uncertainties were estimated from the difference between the ratio
derived from the fitting and the input intrinsic ratio.
The uncertainties of the [\ion{N}{2}]/H$\alpha$ and [\ion{S}{2}]/H$\alpha$
maps were smaller than 30\% and 10\%, respectively.
The uncertainty of the absolute flux calibration was less than 20\%, and the
uncertainty of the H$\alpha$ flux, including the line fitting, was 30\%.
As for the [\ion{S}{2}]$\lambda$6716/[\ion{S}{2}]$\lambda$6731 ratio, it can be
largely affected by the velocity field.
Therefore, the [\ion{S}{2}] ratio map is not presented.
There was a distortion aberration, as expected, in the frames and we have 
corrected it.
The background noise level was $1.9 \times 10^{-16} \ \rm {erg \ cm^{-2}
\ s^{-1} \ arcsec^{-2}}$ in the H$\alpha$ set and $6.3 \times 10^{-17} \
\rm {erg \ cm^{-2} \ s^{-1} \ arcsec^{-2}}$ in the [\ion{S}{2}] set,
respectively.
We determined the absolute coordinate by comparing our FP continuum frame
with the continuum frame of {\it HST} Wide-Field Planetary Camera 2 (WFPC2)
\citep[F675W:][]{Watson:1996}.
The uncertainty of the coordinate between the FP and the {\it HST} frames was
estimated to be 0\arcsec.08 (1$\sigma$).
We adopt a distance of 3.5 Mpc to NGC 253 \citep{Rekola:2005}.

\section{RESULTS}
\label{sec:results}

\subsection{The continuum and H$\alpha$ images}
\label{sec:image-flux}

Figure \ref{fig:panel-center} (a) shows the continuum image of central region
of NGC 253.
Figures \ref{fig:panel-whole} (a), \ref{fig:panel-center} (b), and
\ref{fig:panel-center} (c) show the whole field of view, the central region
(deep), and the central region (shallow) H$\alpha$ images, respectively.
The continuum image is obtained from the average flux of the two off-band
frames.
In the central region, two bright regions are found in the H$\alpha$ and
continuum emission.
One is found at 4\arcsec southwest from the center of NGC 253, while the other
is found at 5\arcsec southeast from the center.
Hereafter, we call the former starburst region A, and the latter starburst
region B.
A filamentary structure is seen to the northeast of the center and we call
it filament C.
In regions farther from the center, there are two bright regions in the
H$\alpha$ image.
One is found at $\sim$30\arcsec east from the center, and the other is found at 
$\sim$60\arcsec southwest from the center. 
We call the former \ion{H}{2} region A, and the latter \ion{H}{2} region B.
We analyse the characteristics of these regions in Section \ref
{sec:line-ratio}.

In the H$\alpha$ and [\ion{N}{2}] images, two faint filaments are seen to the
southeast of the center.
They were found in the narrowband H$\alpha$+[\ion{N}{2}] image (\citealt
{McCarthy:1987}: the H$\alpha$ and [\ion{N}{2}] lines are blended in their
image).
These filamentary structures are observed at the edges of the outflow when
we observe the conical shell-like optically-thin gas (e.g., \citealt
{Strickland:2000, Sugai:2003}).
The positions of these filamentary structures coincide with those in the
near-infrared H$_2$ emission \citep[features A and B:][]{Sugai:2003}.
Hereafter, we call the southwest one filament A and the southeast one
filament B.
We refer to these filamentary structures as "filaments", although we consider
that they are not filaments but actually are a part of conical outflow.
The north-side features were also found in the H$_2$ emission, but not in the
H$\alpha$ probably due to strong extinction.
The positions of filamentary structures seen in the H$\alpha$ and in the
soft X-ray \citep{Strickland:2000} in the galactic wind are close to each
other (Figures \ref{fig:panel-center} (h) and \ref{fig:panel-whole} (b)).
This indicates that the relatively warm ($\sim$ 10$^4$ K) gas exists closely
to the hot ($\sim$ 10$^6$ K) gas.
A closer inspection, however, reveals that the X-ray filaments tend to exist
where the H$\alpha$ emissions are locally weak and tend to exist in inner
regions compared with the H$\alpha$ filaments: the separation between the
H$\alpha$ and the X-ray filaments is $\sim$ 70 pc.
This represents that the inner part of the galactic wind has a higher
temperature than the outer part of it.
This relation between the spatial distributions of the H$\alpha$ and the
X-ray has been found through our accurate positioning of these two maps
(cf. \citealt{Strickland:2000}).
The spatial resolution of our H$\alpha$ image (=1\arcsec.0) also is slightly
better than that obtained by \citealt{Strickland:2000} (=1\arcsec.5).

Starburst region B is as bright as starburst region A in the H$\alpha$ and
the continuum at $\sim$ 6600 \AA, while the counterpart is not found in
the near-infrared \citep{Sams:1994} or radio observations \citep
{Ulvestad:1997}.
This region is diffuse in the optical broadband images retrieved from 
{\it HST} archives.
A comparison of our H$\alpha$ map and long-slit spectroscopic data \citep
{Schulz:1992} has revealed that starburst region B has the sharp local-minimum
velocity of -100 km s$^{-1}$.

\subsection{Line ratio maps}
\label{sec:line-ratio}

\begin{figure}
\epsscale{.80}
\plotone{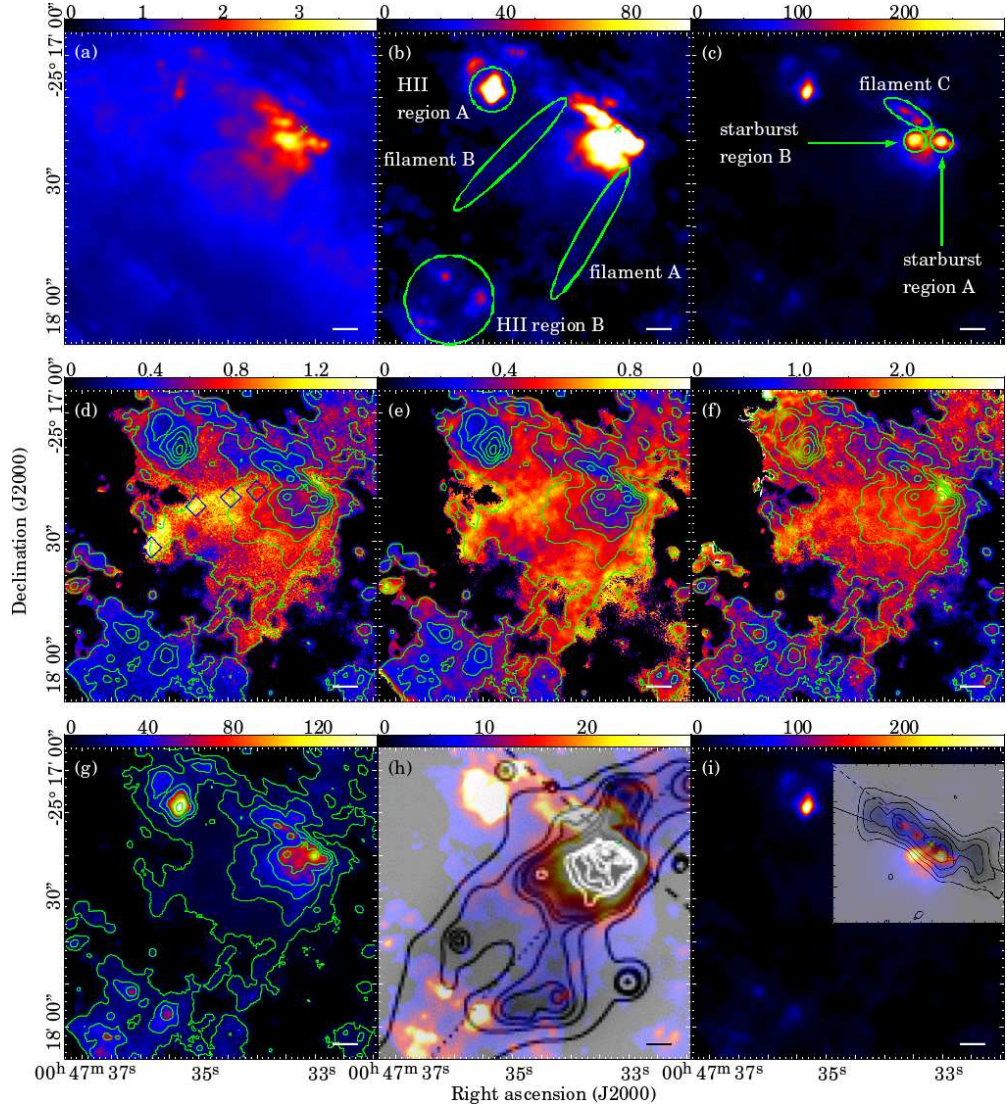}
\caption{The continuum and H$\alpha$ image, and the line ratio maps of the
central region of NGC 253.
North is up and east is to the left.
The green cross at the center of each figure represents the center of NGC 253
derived from the radio continuum \citep{Ulvestad:1997}.
The bars in all figures at lower left represent the length of 100 pc.
(a) The continuum image in units of 10$^{-16}$ erg cm$^{-2}$ s$^{-1}$
arcsec$^{-2}$ \AA$^{-1}$.
(b) The deep and (c) shallow H$\alpha$ image in units of 10$^{-16}$ erg
cm$^{-2}$ s$^{-1}$ arcsec$^{-2}$.
(d) The [\ion{N}{2}]/H$\alpha$ line ratio map.
The background noise level in the H$\alpha$ flux (1$\sigma$) is 1.9 $\times$
10$^{-16}$ erg cm$^{-2}$ s$^{-1}$ arcsec$^{-1}$.
The green contours are 4$\sigma$, 8$\sigma$, 16$\sigma$, 32$\sigma$,
64$\sigma$, and 128$\sigma$ of the H$\alpha$ flux.
The region where the H$\alpha$ flux is weaker than 3$\sigma$ (= 5.7 $\times$
10$^{-16}$ erg cm$^{-2}$ s$^{-1}$ arcsec$^{-2}$) is shown in black.
The blue squares represent the regions where the [\ion{N}{2}]/H$\alpha$ and 
[\ion{S}{2}]/H$\alpha$ ratios are measured and are plotted with green squares
in Figure \ref{fig:shock-n2-s2}.
(e) The [\ion{S}{2}]/H$\alpha$ line ratio map.
(f) The [\ion{N}{2}]/[\ion{S}{2}] line ratio map.
(g) The H$\alpha$ equivalent width map in units of \AA.
(h) The continuum subtracted H$\alpha$ image in units of 10$^{-16}$ erg
cm$^{-2}$ s$^{-1}$ arcsec$^{-2}$.
The contours show the intensity of soft X-ray \citep[0.3-2.0 keV:][]
{Strickland:2000}.
(i) The continuum subtracted H$\alpha$ image in units of 10$^{-16}$ erg
cm$^{-2}$ s$^{-1}$ arcsec$^{-2}$.
The contours show the intensity of CO J = 1 $\rightarrow$ 0 \citep
{Paglione:2004}.
\label{fig:panel-center}}
\end{figure}

\begin{figure}
\epsscale{1.00}
\plotone{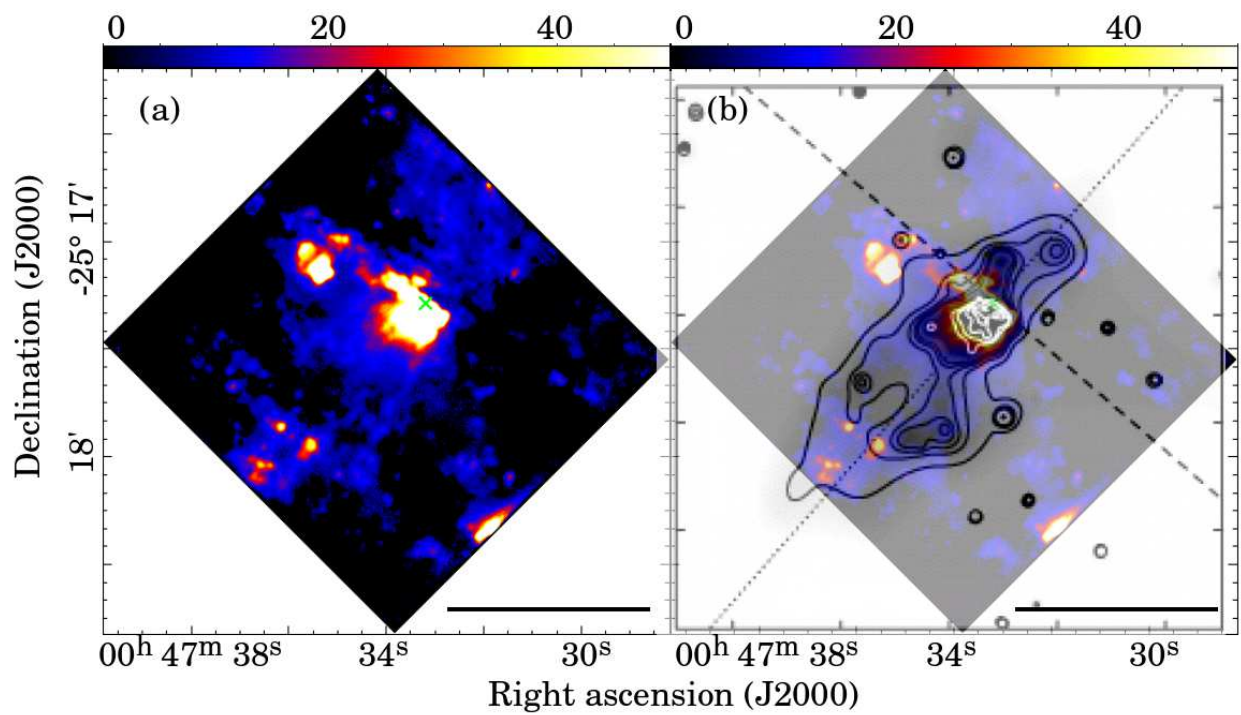}
\caption{(a) The continuum subtracted H$\alpha$ images of NGC 253 in units of
erg cm$^{-2}$ s$^{-1}$ arcsec$^{-2}$.
North is up and east is to the left.
The green cross at the center of this figure represents the center of NGC 253
derived from the radio continuum \citep{Ulvestad:1997}.
The bar at lower left represents the length of 1 kpc.
The region where the H$\alpha$ flux is weaker than 3$\sigma$ (= 5.7 $\times$
10$^{-16}$ erg cm$^{-2}$ s$^{-1}$ arcsec$^{-2}$) is shown in black.
The bright point at the lower right edge is a ghost of the central region. 
(b) The continuum subtracted H$\alpha$ images of NGC 253 in units of
erg cm$^{-2}$ s$^{-1}$ arcsec$^{-2}$.
The contours show the intensity of soft X-ray \citep[0.3-2.0 keV:][]
{Strickland:2000}.
\label{fig:panel-whole}}
\end{figure}

Figures \ref{fig:panel-center} (d) and \ref{fig:panel-center} (e) show the 
[\ion{N}{2}]$\lambda$6583/H$\alpha$ and
[\ion{S}{2}]$\lambda\lambda$6716,6731/H$\alpha$ ratio maps, respectively.
The [\ion{N}{2}]/H$\alpha$ line ratio varies widely among regions.
The [\ion{N}{2}]/H$\alpha$ ratios of filaments A and B are larger than the
other regions.
The ratio is particularly the largest at the tip of filament B and reaches
about 1.5.
As leaving from the galactic center along the minor axis, the H$\alpha$ flux
of the galactic wind becomes weaker while its [\ion{N}{2}]/H$\alpha$ ratio
tends to become larger.
The [\ion{N}{2}]/H$\alpha$ ratios of starburst regions A and B are $\sim$ 0.5,
which are local minima in the central region and the galactic wind.
The [\ion{N}{2}]/H$\alpha$ ratios of \ion{H}{2} regions A and B are $\sim$ 0.25
and are smaller than in the other regions.
These ratios coincide with those of \ion{H}{2} regions in nearby disk galaxies
\citep{Moustakas:2006}.
Therefore, we conclude that these regions are \ion{H}{2} regions, which is why
we call them \ion{H}{2} regions A and B.
As for \ion{H}{2} region B, \citet{McCarthy:1987} observed a part of this
region and suggested that it is an \ion{H}{2} region.
Our data show fairly uniform distribution of the [\ion{N}{2}]/H$\alpha$ ratio
of $\sim$ 0.25 over the whole \ion{H}{2} region B including its three clumps.
The [\ion{N}{2}]/H$\alpha$ ratio of filament C is similar to those of
\ion{H}{2} regions A and B, rather than those of the galactic wind or
starburst regions A and B.
Therefore, this region is a star-forming region located on the same line of
sight as an edge of the galactic wind although the region looks as if
it is a part of the galactic wind.
In Figure \ref{fig:panel-center} (i) we compare our H$\alpha$ map with the CO
intensity map \citep{Paglione:2004}.
We find that the position of filament C coincides with that of a strong CO
emission elongated along the $x_2$ orbit \citep{Binney:1991,
Athanassoula:1992}, which is made by and is oriented perpendicular to the bar
\citep{Peng:1996, Paglione:2004}.
Hence, star formation in filament C occurs due to this $x_2$ orbit.
The [\ion{S}{2}]$\lambda\lambda$6716,6731/H$\alpha$ ratio map is globally
similar to the [\ion{N}{2}]/H$\alpha$ ratio map: e.g., the
[\ion{S}{2}]/H$\alpha$ ratios of filaments A and B are the largest among
regions while those of \ion{H}{2} regions A and B are the smallest.
The [\ion{S}{2}]/H$\alpha$ ratios remain large outside of the filaments while
the [\ion{N}{2}]/H$\alpha$ ratios become smaller.
This may be because the metallicity of the galactic winds is larger than
that outside of the filaments (see Section \ref{sec:N-enrich}).

\begin{figure}
\epsscale{.80}
\plotone{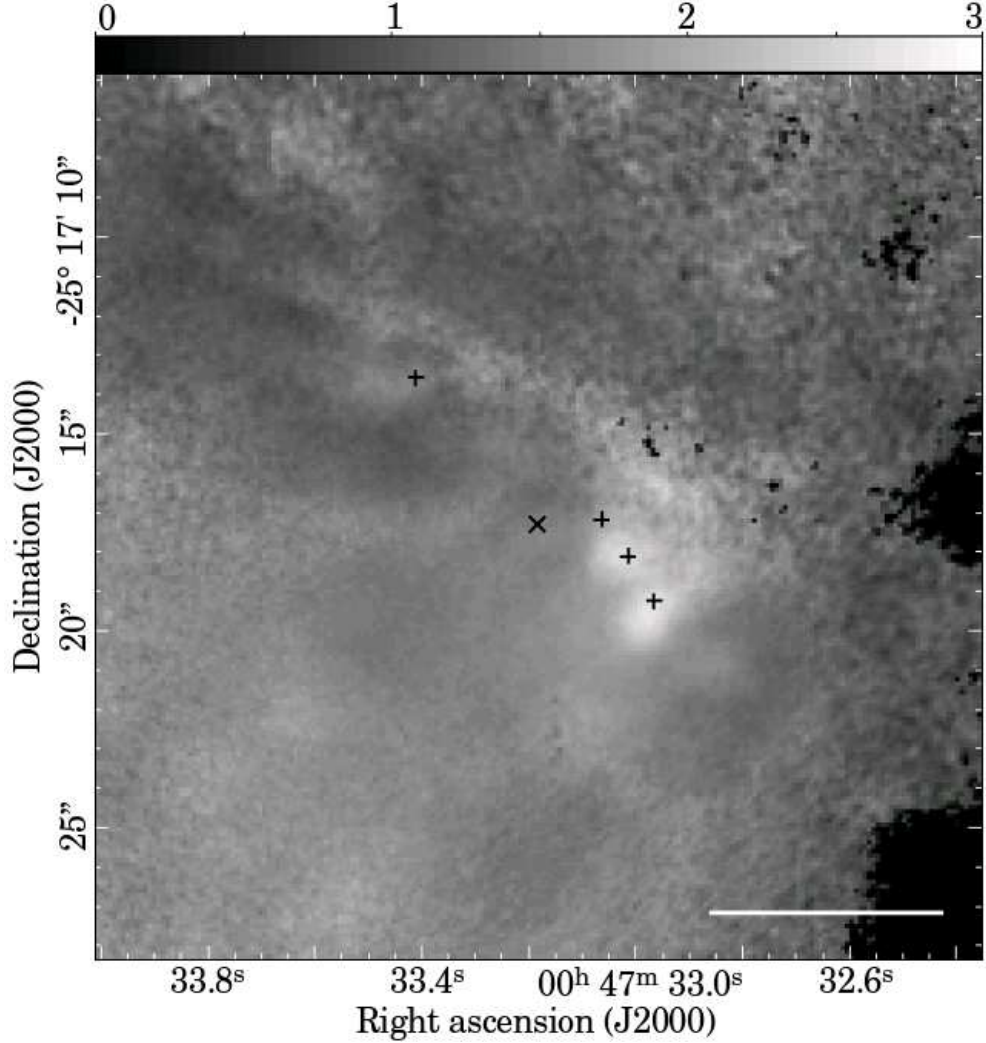}
\caption{The [\ion{N}{2}]$\lambda$6583/[\ion{S}{2}]$\lambda\lambda$6716,6731
ratio map in the central region.
North is up and east is to the left.
The cross at the center of this figure represents the center of NGC 253
derived from the radio continuum \citep{Ulvestad:1997}.
The bar at lower left represents the length of 100 pc.
Four plus signs represent the positions of star clusters found by {\it HST},
which are spot a, spot n, spot i, and the bright blob starting from the left.
The region where the H$\alpha$ flux is weaker than 3$\sigma$ (= 5.7 $\times$
10$^{-16}$ erg cm$^{-2}$ s$^{-1}$ arcsec$^{-2}$) is shown in black.
[{\it See the electronic edition of the Journal for a color version of this
figure.}]
\label{fig:NII-SII-bb}}
\end{figure}

Figure \ref{fig:panel-center} (f) shows the 
[\ion{N}{2}]$\lambda$6583/[\ion{S}{2}]$\lambda\lambda$6716,6731 ratio map.
The [\ion{N}{2}]/[\ion{S}{2}] ratios in the central region and the galactic
wind are almost uniform and are $\sim$ 1.5, while the H$\alpha$ fluxes vary.
The ratio of filament C is small and is $\sim$ 1.0.
Two large [\ion{N}{2}]/[\ion{S}{2}] regions, whose ratios are greater than 2.0,
are found to the north of starburst region A (Figure \ref{fig:NII-SII-bb}).
Their positions coincide with those of star clusters found in {\it HST}
observation \citep[the bright blob and spot i:][]{Watson:1996}.
The cause of the large [\ion{N}{2}]/[\ion{S}{2}] ratio is probably the nitrogen
enrichment due to star clusters.
We will discuss this in detail in Section \ref{sec:N-enrich}.

Figure \ref{fig:panel-center} (g) shows the H$\alpha$ equivalent width map
calculated from the H$\alpha$ and continuum images.
The equivalent width of \ion{H}{2} region A is 95 \AA~on average and larger
than 150 \AA~at its center, which is the largest among regions.
Starburst regions A and B have local maxima of the equivalent width in the
central region: 88 \AA~and 64 \AA, respectively.
Three clumps of \ion{H}{2} region B have about the same equivalent widths with
each other, ranging from 36 \AA~to 39 \AA.
When the underlying diffuse continuum emission is subtracted, the equivalent
widths of \ion{H}{2} regions A and B are $\sim$ 300--500 \AA~(Table \ref
{tb:flux-ratio}). 
As diffuse continuum emission regions, we selected regions that are near the
\ion{H}{2} regions and do not include extra emission.
This estimation of the H$\alpha$ equivalent widths has an uncertainty of a
factor 2 at most, mainly caused by the uncertainty of determination of this
diffuse continuum level.

The aperture, the H$\alpha$ flux, the line ratios, and the equivalent width
of each region are listed in Table \ref{tb:flux-ratio}.
The star formation rate is estimated from the H$\alpha$ flux and also is
listed in the table.
As an example, we adopted the 6\arcsec.0 diameter circular aperture in measuring
values of \ion{H}{2} region A.
The measured H$\alpha$ flux of \ion{H}{2} region A is 3.5$\times 10^{-13}$ erg
cm$^{-2}$ s$^{-1}$, and its H$\alpha$ luminosity is 5.1$\times 10^{38}$
erg s$^{-1}$.
From this H$\alpha$ luminosity, the number of ionizing photons is
estimated to be 3.7$\times 10^{50}$ s$^{-1}$ and the star formation rate is
estimated to be 4.0$\times 10^{-3}$ $M_\odot$ yr$^{-1}$ \citep
{Kennicutt:1998}.

\section{DISCUSSION}
\label{sec:discuss}

\subsection{Filament B}
\label{sec:filament-B}

At the tip of filament B, the [\ion{N}{2}]/H$\alpha$ ratio reaches about 1.5
(see Section \ref{sec:line-ratio}).
An ordinary photoionization process in \ion{H}{2} regions cannot produce this
large [\ion{N}{2}]/H$\alpha$ ratio of filament B, while shock
ionization/excitation is plausible.
Although \citet{Schulz:1992} produced the [\ion{N}{2}]/H$\alpha$ ratio larger
than 1.0 with photoionization model, this model used a single stellar
temperature for ionizing stars.
The photoionization model by \citet{Kewley:2001} is more reliable, because
they took into consideration stellar population.
\citet{Kewley:2001} placed a theoretical upper limit for starburst models 
on the optical diagnostic diagrams (their equation (5) and Figure 16(a)),
which show that it is difficult to exceed 1.0 in the [\ion{N}{2}]/H$\alpha$
ratio with starbursts even though they took into consideration the effects of
the nitrogen enrichment.
No combination of parameters, such as star formation history and ionization
parameter, can generate a theoretical point above this fold.
This limit is rather conservative, and \citet{Kauffmann:2003} placed a
demarcation line at lower [\ion{N}{2}]/H$\alpha$ ratio from observational data.
Photoionization may somewhat contribute near the central region, where the
[\ion{N}{2}]/H$\alpha$ ratio is smaller than 1.0.

\begin{figure}
\epsscale{.80}
\plotone{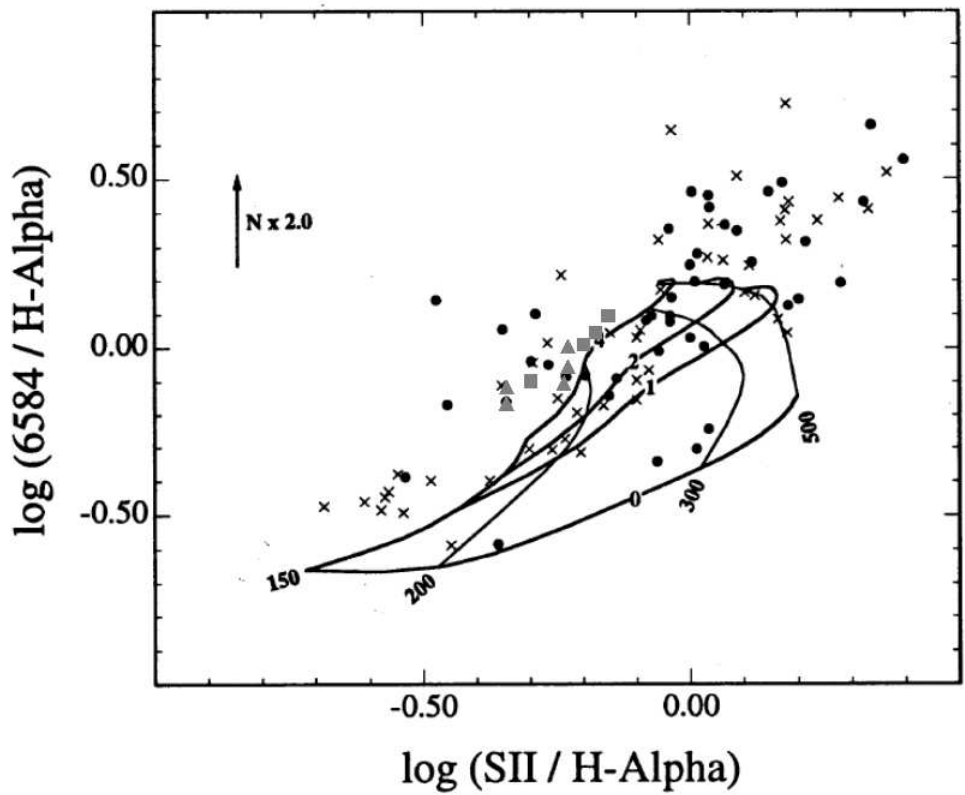}
\caption{The [\ion{N}{2}]/H$\alpha$ ratio vs the [\ion{S}{2}]/H$\alpha$ ratio
diagram of the shock model \citep[shock-only:][]{Dopita:1995}.
The numbers below the model curves represent the shock velocities in units of
km s$^{-1}$, and the numbers on the curves represent the magnetic parameters
($B/n^{1/2}$) in units of $\mu$G cm$^{3/2}$.
The effect of a factor of 2 increase in the N/O ratio is shown by the arrow.
Filled circles are for LINERs while crosses are for radio galaxies \citep
{Dopita:1995}.
Four squares represent the line ratios observed at filament B in NGC 253 (see
Figure \ref{fig:panel-center} (d) for their positions) while five triangles
represent those in typical regions of the galactic wind.
The square drawn furthest left shows the line ratios closest to the galactic
disk, with each subsequent square to the right lying at increasing distance.
[{\it See the electronic edition of the Journal for a color version of this
figure.}]
\label{fig:shock-n2-s2}}
\end{figure}

According to the shock model calculated by \citet{Dopita:1995}, the
[\ion{N}{2}]/H$\alpha$ ratio reaches to 1.55, in the case of the
solar-abundance interstellar gas, the magnetic parameter $B/n^{1/2}$ of 2
$\mu$G cm$^{3/2}$, which is the intermediate value among their models, and
shock velocity of 500 km s$^{-1}$.
This model can explain the largest [\ion{N}{2}]/H$\alpha$ ratio in our data,
$\sim$ 1.5.
By using the [\ion{S}{2}]$\lambda\lambda$6716,6731/H$\alpha$ ratio map and the
[\ion{N}{2}]/H$\alpha$ ratio map, it is possible to discuss the possibility of
nitrogen enrichment.
Figure \ref{fig:shock-n2-s2} shows the [\ion{N}{2}]/H$\alpha$ ratio vs the
[\ion{S}{2}]/H$\alpha$ ratio diagram of a shock model \citep[shock-only:][]
{Dopita:1995}.
The observed ratios (squares and triangles) plotted above the model curves
indicate that the N/O ratio is about 1.8 times the solar value when we assume
that the magnetic parameter is equal to 2 and there is no contribution of
photoionization for these ratios.
We used the \citet{Anders:1989} abundance set, because photoionization and
shock models that we referred to \citep{Kewley:2001, Dopita:1995} use it.
If new abundance set \citep{Grevesse:2007} is used and if the model flux is
proportional to the amount of each element, the N/O ratio will be 2.5 times
the solar value.
Both the [\ion{N}{2}]/H$\alpha$ and the [\ion{S}{2}]/H$\alpha$ ratios are
larger in further positions from the galactic disk in filament B.
This means that the shock velocity or the degree of contribution of
photoionization varies among these positions.

We compare the [\ion{N}{2}]/H$\alpha$ ratio of the galactic wind in NGC 253
with those in other galaxies.
The [\ion{N}{2}]/H$\alpha$ ratio of the galactic wind in M82, which is one of
the most famous edge-on galaxies, is mainly less than 1.0, and photoionization
by starburst is considered to be an ionization/excitation source \citep
{Shopbell:1998}.
There may be an influence of shock since the [\ion{N}{2}]/H$\alpha$ ratio
tends to increase far from the galactic disk.
In NGC 1482, in contrast, the [\ion{N}{2}]/H$\alpha$ ratio is larger than 1.0
in the bulk of the galactic wind and reaches to 2.0 \citep{Veilleux:2002}.
This indicates shock as the ionization/excitation source.
The [\ion{N}{2}]/H$\alpha$ ratio of galactic wind in NGC 253 is between the
ratios observed for galactic winds in M82 and NGC 1482.
The typical [\ion{N}{2}]/H$\alpha$ ratios are 0.3-0.6 in M82
\citep{Shopbell:1998} and 1.0-2.3 in NGC 1482 \citep{Veilleux:2002}, while it
is 0.8-1.5 in NGC 253.

The shape of galactic wind and the line ratio maps are non-axisymmetric
about the galactic minor axis, which is also seen in M82 \citep{Shopbell:1998}.
The non-axisymmetric structure in NGC 253 is also found in the H$_2$ emission,
which suggests the inhomogeneous nature of the interstellar medium \citep
{Sugai:2003}.
Our [\ion{N}{2}]/H$\alpha$ ratio map shows that filament B has a larger ratio
than filament A.
This may occur if filament A is more affected by photoionization compared
with filament B, or if the mass or the velocity of the wind gas varies from
region to region.

\subsection{Nitrogen enrichment}
\label{sec:N-enrich}

Starburst regions A and B emit the H$\alpha$ and the continuum strongly and
have low [\ion{N}{2}]/H$\alpha$ ratios.
Their [\ion{N}{2}]/H$\alpha$ ratios are larger than those of \ion{H}{2} regions
in nearby disk galaxies \citep{Moustakas:2006}, and are about twice as large
as those of \ion{H}{2} regions A and B in NGC 253 itself (Figure \ref
{fig:panel-center} (d)).
These large [\ion{N}{2}]/H$\alpha$ ratios are probably caused by nitrogen
enrichment, as discussed in the following.
Nitrogen enrichment relative to oxygen in metal rich environment was found in
a statistical work by \citet{Liang:2006}.
The nitrogen enrichment can easily explain the observed large
[\ion{N}{2}]/H$\alpha$ as well as [\ion{N}{2}]/[\ion{S}{2}] ratio although
distorted ionizing continuum, which mimicks the situation where neutral matter
lies between the ionizing stars and the emission-line clouds
\citep{Schulz:1992}, or shock may also increase the [\ion{N}{2}]/H$\alpha$
ratio.

Nitrogen enrichment occurs in the following process (e.g., \citealt
{Pagel:1997, Sugai:2004b}).
There are two kinds of elements produced in stars, i.e., the primary element
and the secondary element.
The primary element, such as oxygen and sulfur, is produced from hydrogen and
helium.
The ratio of the produced amount of each element (e.g., sulfur to oxygen) is
constant and insensitive to the metallicity at the star birth.
On the other hand the secondary element, such as nitrogen, is produced also
in the CNO cycle, and is spread into the interstellar medium with the stellar
wind.
The produced amount increases with an increase in the metallicity at the
star birth.
Therefore, in a metal rich environment, the ratio of the produced amount of
nitrogen (secondary element) to that of sulfur (primary element) becomes
larger, and the [\ion{N}{2}]/[\ion{S}{2}] ratio becomes large.
The relationship between the [\ion{N}{2}]/[\ion{S}{2}] ratio and the
metallicity of galaxies is found in SDSS observations \citep{Liang:2006}.

In the [\ion{N}{2}]/[\ion{S}{2}] ratio map (Figure \ref{fig:panel-center} (f)),
the [\ion{N}{2}]/[\ion{S}{2}] ratios of starburst regions A and B are larger
than those of \ion{H}{2} regions A and B and filament C.
This indicates that starburst regions A and B have higher metallicity.
The [\ion{N}{2}]/[\ion{S}{2}] ratios in the bright blob and spot i (hereafter,
we call it blob/spot region) are the largest (Figure \ref{fig:NII-SII-bb}),
so that the degree of nitrogen enrichment is extremely high.
The de-excitation of [\ion{S}{2}] \citep{Osterbrock:2006} may cause this large
[\ion{N}{2}]/[\ion{S}{2}] ratio.
The electron densities are estimated to be no more than $\sim$ 900 cm$^{-3}$
from [\ion{S}{2}]$\lambda$6716/[\ion{S}{2}]$\lambda$6731 ratios observed by
\citet {Schulz:1992}.
This ratio is at the minimum ($\sim$ 0.9) around the blob/spot region
in the whole field of view of their spectroscopic observations.
These electron densities are small and the de-excitation reduces the
[\ion{S}{2}] flux by no more than 16 \%.
The bright blob and spot i are bright in J, H, and K band \citep{Sams:1994},
and are two out of four clusters defined in the optical continuum in {\it HST}
observation \citep{Watson:1996}.
The bright blob is the brightest cluster among the four.
Spot i is the second brightest in the H$\alpha$ flux.
We estimate the metallicity of blob/spot region from the
[\ion{N}{2}]/[\ion{S}{2}] ratio, 2.19.
This ratio is at the largest end of the distribution for starburst galaxies
analyzed by \citet{Liang:2006}.
Assuming the relation between the [\ion{N}{2}]/[\ion{S}{2}] ratio and the
metallicity in starburst galaxies \citep[Figure 3(d) by][]{Liang:2006}, the
metallicity (12+log(O/H)) of blob/spot region is derived as 9.17 $\pm$ 0.08
(1$\sigma$).
Using the relation between log(O/H) and log(N/O) (Figure 8 by
\citealt{Liang:2006}), the log(N/O) abundance of blob/spot region is estimated
to be -0.50 $\pm$ 0.13 (1$\sigma$).
Here the scatter in N/O to O/H in Figure 8 by \citet{Liang:2006} was estimated
to be about 0.1 dex (1$\sigma$) by eye-inspection.
When we use the \citet{Anders:1989} abundance set, whose log(N/O) ratio is
-0.88, the N/O ratio of blob/spot region is about 2.4 times larger than the
solar value.
This value of 2.4 does not change even if we use the \citet{Grevesse:2007}
abundance set, because its N/O ratio is equal to that of the
\citet{Anders:1989} abundance set.
This value is similar to that in the galactic wind (Section \ref
{sec:filament-B}), although it is difficult to estimate the abundance
precisely.
To explain this similarity, it is plausible that the gas of the galactic wind
is chemically evolved in starburst regions and then is blown out.

\subsection{Extent of the galactic wind}

In \ion{H}{2} region B, three clumps are found in our H$\alpha$ map.
The northern and western clumps are on the spiral arm, while the southeastern
one is not.
Soft X-ray (0.3-2.0 keV) is detected around \ion{H}{2} region B, including the
three clumps (Figures \ref{fig:panel-center} (h) and \ref{fig:panel-whole} (b)).
Therefore, it is difficult to determine from only the H$\alpha$ map whether
the southeastern clump is a part of the galactic wind or an \ion{H}{2} region
on the galactic disk with an X-ray emitting gas accidentally located on the
same line of sight.
This issue is important when considering the structure of the galactic wind,
i.e., whether or not the tip of the galactic wind is cooled down enough to
emit the H$\alpha$.
Although \ion{H}{2} region B was observed and considered to be \ion{H}{2}
regions through a long-slit spectroscopy \citep{McCarthy:1987}, the line ratio
of each clump was not discussed.
In our data, the southeastern clump, as well as the northern and the western
ones, have the same [\ion{N}{2}]/H$\alpha$ and [\ion{S}{2}]/H$\alpha$ ratios
as \ion{H}{2} regions in nearby disk galaxies \citep{Moustakas:2006}.
Moreover, the equivalent widths of three clumps are similar: $\sim$ 40
\AA~(Figure \ref{fig:panel-center} (g)).
Consequently, we consider that the three clumps of \ion{H}{2} region B are not
a part of the galactic wind but are \ion{H}{2} regions on the galactic disk.

\subsection{Kinetic energy of the galactic wind}

We derive the properties of the galactic wind, given that the region where
the [\ion{N}{2}]/H$\alpha$ ratio is greater than 1.0 (or the
[\ion{S}{2}]/H$\alpha$ ratio is greater than 0.5) is the galactic wind
\citep{Veilleux:2005, Bland:2007}.
First we estimate the H$\alpha$ luminosity of the wind.
When the [\ion{N}{2}]/H$\alpha$ criterion is used, the H$\alpha$ flux of the
galactic wind is equal to 1.6$\times 10^{-13}$ erg cm$^{-2}$ s$^{-1}$, and
its H$\alpha$ luminosity is 2.3$\times 10^{38}$ erg s$^{-1}$ when 3.5 Mpc
is used as a distance to NGC 253 \citep{Rekola:2005}.
When the [\ion{S}{2}]/H$\alpha$ criterion is used instead, the H$\alpha$ flux
is equal to 8.9$\times 10^{-13}$ erg cm$^{-2}$ s$^{-1}$ and the H$\alpha$
luminosity is 1.3$\times 10^{39}$ erg s$^{-1}$.
Hereafter, we adopt the [\ion{S}{2}]/H$\alpha$ criterion, because it is less
affected by nitrogen enrichment (Section \ref{sec:filament-B},
\ref{sec:N-enrich}) compared with the [\ion{N}{2}]/H$\alpha$ criterion.
Because the total H$\alpha$+[\ion{N}{2}]$\lambda\lambda$6548,6583 luminosity
of NGC 253 is equal to 1.1$\times 10^{41}$ erg s$^{-1}$ \citep{Hoopes:1996},
the H$\alpha$ luminosity of the galactic wind is only from 0.5\% to 5\% of
the total luminosity.
The actual fraction of the luminosity of the galactic wind may be even
smaller, because extinction is not corrected in the above estimate although
the central region of NGC 253 is heavily dusty (e.g., \citealt{Sams:1994}).
Secondly, we estimate the mass of ionized gas in the galactic wind.
From the [\ion{S}{2}]$\lambda$6716/[\ion{S}{2}]$\lambda$6731 ratio of 0.9 to
1.2 in the central region \citep{Schulz:1992}, the electron density is
estimated to be from 100 cm$^{-3}$ to 1000 cm$^{-3}$ \citep{Osterbrock:2006}.
Using the electron density of 100$n_{e,2}$ cm$^{-3}$, where $n_{e,2}$ is
normalized to 100 cm$^{-3}$, and the recombination rate of H$\alpha$ of
8.7$\times 10^{-14}$ cm$^3$ s$^{-1}$ \citep{Osterbrock:2006}, the H$\alpha$
emitting gas mass of the galactic wind is estimated to be 8.6$\times 10^4
n_{e,2}^{-1}~M_{\odot}$.
We assumed that the northwest side of the galactic wind, which is not detected
in our observations due to heavy extinction, actually exists as well as the
southeast side of it (Section \ref{sec:image-flux}).
For simplicity, we assumed the symmetric shape of the galactic wind in order
to roughly estimate the volume filling factor, although it is non-axisymmetric
about the galactic minor axis.
The assumed shape of the galactic wind is a truncated cone structure whose
opening angle is 30$^\circ$ and diameter of the base and slant height are 500
pc and 700 pc, respectively.
These values are measured through the H$\alpha$ map.
Then, the volume filling factor of the galactic wind is estimated to be 4.2
$\times$ 10$^{-4}$ $n_{e,2}^{-2}$.
The limb-brightening in the H$\alpha$ and the X-ray (Section
\ref{sec:image-flux}) and the small filling factor suggest that the galactic
wind has a hollow structure.
Finally, using the outflow velocity of 390 km s$^{-1}$ \citep{Schulz:1992},
the kinetic energy of the galactic wind is estimated to be 1.3$\times 10^{53}
n_{e,2}^{-1}$ erg.
The dynamical time scale $\tau_{dyn}$ derived from the length of the
filament B ($\sim$ 700pc) and the outflow velocity (390 km s$^{-1}$) is $\sim$
1.8 Myr, and the kinetic energy is increased by 2.3$\times 10^{39}
n_{e,2}^{-1}$ erg s$^{-1}$ on average.

In order to consider whether the energy input from supernovae can supply
the kinetic energy of the galactic wind, we estimate first the H$\alpha$
luminosity of the bright blob, which is the brightest region in the central
region of NGC 253.
Because of heavy extinction in the central region of NGC 253, we have
corrected the effect of extinction using the ratio of the H$\alpha$ flux to
the Br$\gamma$ flux \citep{Forbes:1993}.
We assumed the intrinsic Br$\gamma$/H$\alpha$ flux ratio of 9.80$\times
10^{-3}$ \citep{Osterbrock:2006} and the extinction law of A(6563\AA)=0.79A$_V$
and A(2.2$\mu$m)=0.15A$_V$ \citep{Whitford:1958, Miller:1972}.
The H$\alpha$ flux of the bright blob in 2\arcsec aperture, which is the same
as used by \citet{Forbes:1993}, is 5.5$\times 10^{-14}$ erg cm$^{-2}$ s$^{-1}$,
and A$_V$ is estimated to 6.9 mag.
The extinction-corrected H$\alpha$ flux is 8.3$\times 10^{-12}$ erg cm$^{-2}$
s$^{-1}$, and the H$\alpha$ luminosity is 1.2$\times 10^{40}$ erg s$^{-1}$,
which corresponds to the star formation rate of 0.096 $M_{\odot}$ yr$^{-1}$
\citep{Kennicutt:1998}.
When we assume that the star formation rate of the bright blob is constant,
the supplied kinetic energy through supernovae and stellar winds from this
region in 1.8 Myr ($=\tau_{dyn}$) is estimated to be 5.6$\times 10^{53}$ erg
\citep[Starburst99:][]{Leitherer:1999}.
We assumed Salpeter initial mass function \citep{Salpeter:1955},
$M_{up}$=100$M_{\odot}$ and $M_{low}$=1$M_{\odot}$ as cut-off masses,
metallicity Z=0.020, and that the galactic wind began to blow when the first
supernova occurred in the bright blob (3.6 Myr: Figure 114 in
\citealt{Leitherer:1999}).
Therefore, the energy input from the bright blob is much greater than the
kinetic energy of the galactic wind.
This suggests that starbursts are the main energy source of the galactic
wind in NGC 253.

We compare the properties of the galactic winds in M82 \citep{Shopbell:1998},
NGC 1482 \citep{Veilleux:2002}, and NGC 253, although the definition of
galactic wind regions in M82 is quite different from that in NGC 253 and NGC
1482.
\citet{Shopbell:1998} defined the galactic wind regions by simply excluding
within approximately 8\arcsec of the disk (+ halo component), while we and
\citet{Veilleux:2002} used the criterion of the [\ion{S}{2}]/H$\alpha$ and
[\ion{N}{2}]/H$\alpha$ ratio, respectively.
The kinetic energy of the galactic wind in NGC 253 (1.3$\times 10^{53}$ erg)
is found to be smaller than those in M82 (2.1$\times 10^{55}$ erg) and in
NGC 1482 (2$\times 10^{53}$ erg).
There are three possibilities for the difference in the energies of the
galactic winds: the masses, the velocities, and the dynamical timescales.
The mass in NGC 1482 is about 4.2 times, and the mass in M82 is about 67 times
as large as that in NGC 253 (8.6$\times 10^4 M_{\odot}$).
The outflow velocities and the dynamical timescales do not vary drastically
among these galaxies: 390 km s$^{-1}$ and 1.8 Myr in NGC 253, 250 km s$^{-1}$
and 6 Myr in NGC 1482, and 600 km s$^{-1}$ and 3.4 Myr in M82, respectively.
Therefore, the difference of the outflow mass in each galaxy, which may depend
on the strength of starburst in the central region, probably dominantly makes
the variety of the kinetic energy of the galactic wind, although the
combination of the differences of the mass, the velocity, and the dynamical
time scale will determine the exact value.

\section{SUMMARY}
\label{sec:summary}

We have observed the central region of NGC 253 with the Fabry-Perot mode of 
the Kyoto3DII mounted on the Subaru Telescope.
We have presented the continuum and the H$\alpha$ images, and the
[\ion{N}{2}]$\lambda$6583/H$\alpha$, 
[\ion{S}{2}]$\lambda\lambda$6716,6731/H$\alpha$, and
[\ion{N}{2}]$\lambda$6583/[\ion{S}{2}]$\lambda\lambda$6716,6731 ratio maps.
The [\ion{N}{2}]/H$\alpha$ of filament B is larger than 1.0, which indicates
that this region is ionized/excited by shock.
Non-axisymmetry in the galactic wind, which is also seen in M82, is found.
The shapes of the galactic wind and the line ratio maps are non-axisymmetric
about the galactic minor axis: e.g., the [\ion{N}{2}]/H$\alpha$ of filament A
is smaller than that of filament B.
The [\ion{N}{2}]/H$\alpha$ map enables us to isolate filament C from the
galactic wind.
Using the [\ion{S}{2}]/H$\alpha$ map, we extract the galactic wind region, and
estimate the mass and the kinetic energy of the galactic wind: 8.6$\times 10^4
M_{\odot}$ and 1.3$\times 10^{53}$ erg, respectively.
The mass and the kinetic energy of the galactic wind in NGC 253 are smaller
than those in M82 and NGC 1482.
The filling factor of the galactic wind is much smaller than unity, which is
consistent with that the galactic wind has a hollow structure.
The energy input from supernovae in the bright blob can sufficiently supply
the energy of the galactic wind.
From the [\ion{N}{2}]/[\ion{S}{2}] ratio map, the interstellar gas in the
central region and in the galactic wind is found to have a high nitrogen
enrichment.
The bright blob and spot i, which are stellar clusters found in {\it HST}
observations, have especially high [\ion{N}{2}]/[\ion{S}{2}] ratios due to
heavy nitrogen enrichment.

\acknowledgments

We thank the staff at the Subaru Telescope for their help during the test 
observing run.
We also thank T. Ohtani, T. Hayashi, M. Ishii, Y. Okita, A. Akita, J. Sudo,
M. Sasaki, N. Takeyama, M. A. Malkan, and A. R. Jenner for discussions.
This work was supported by the Grant-in-Aid for the Global COE Program "The
Next Generation of Physics, Spun from Universality and Emergence" from the
Ministry of Education, Culture, Sports, Science and Technology (MEXT) of Japan.

\clearpage

\clearpage

\clearpage

\clearpage

\clearpage

\clearpage

\begin{deluxetable}{cccccccc}
\rotate
\tablecaption{Fluxes, star formation rates, and line ratios of each region
\label{tb:flux-ratio}}
\tablewidth{0pt}
\tablehead{
\colhead{Region} & \colhead{Aperture} & \colhead{Flux(H$\alpha$)\tablenotemark
{a}} & \colhead{SFR\tablenotemark{b}} &
\colhead{EW(H$\alpha$)\tablenotemark{c}} & 
\colhead{[\ion{N}{2}]/H$\alpha$\tablenotemark{d}} &
\colhead{[\ion{S}{2}]/H$\alpha$\tablenotemark{e}} &
\colhead{[\ion{N}{2}]/[\ion{S}{2}]\tablenotemark{d,}\tablenotemark{e}}
}
\startdata
starburst region A & 3".5 & 1.9 & 2.2 & 88 & 0.54 & 0.28 & 1.93 \\
starburst region B & 4".0 & 2.5 & 2.9 & 64 & 0.51 & 0.35 & 1.45 \\
filament C & 11".54 $\times$ 3".47 & 3.0 & 3.5 & 42 & 0.38 & 0.32 & 1.16 \\
\ion{H}{2} region A & 6".0 & 3.5 & 4.0 & 95(482) & 0.25 & 0.17 & 1.45 \\
\ion{H}{2} region B(N) & 4".0 & 0.34 & 0.39 & 36(316) & 0.24 & 0.22 & 1.06 \\
\ion{H}{2} region B(W) & 6".0 & 0.72 & 0.83 & 36(371) & 0.26 & 0.24 & 1.09 \\
\ion{H}{2} region B(SE) & 8".0 & 1.1 & 1.3 & 39(282) & 0.26 & 0.26 & 0.98 \\
\enddata
\tablecomments{The errors for flux(H$\alpha$), EW(H$\alpha$),
[\ion{N}{2}]/H$\alpha$, [\ion{S}{2}]/H$\alpha$, and [\ion{N}{2}]/[\ion{S}{2}] 
are less than 30\%, a factor of 2, 30\%, 10\%, and 30\%, respectively.}
\tablenotetext{a}{The unit of flux is 10$^{-13}$ erg cm$^{-2}$ s$^{-1}$.}
\tablenotetext{b}{The star formation rates are estimated from the H$\alpha$
fluxes using the relation obtained by \citet{Kennicutt:1998}.
The unit of star formation rate is 10$^{-3}$ $M_{\odot}$ yr$^{-1}$.}
\tablenotetext{c}{The unit of equivalent width is \AA.
The value shown in parentheses is the equivalent width derived by subtracting
the underlying diffuse continuum from the continuum emission.}
\tablenotetext{d}{[\ion{N}{2}] means the [\ion{N}{2}]$\lambda$6583 line.}
\tablenotetext{e}{[\ion{S}{2}] means the [\ion{S}{2}]$\lambda\lambda$6716,6731
lines.}
\end{deluxetable}


\begin{thebibliography}{}

\bibitem[Anders \& Grevesse(1989)]{Anders:1989} Anders, E., \& Grevesse,
    N.\ 1989, \gca, 53, 197
\bibitem[Athanassoula(1992)]{Athanassoula:1992} Athanassoula, E.\ 1992, 
    \mnras, 259, 328
\bibitem[Binney et al.(1991)]{Binney:1991} Binney, J., Gerhard, O.~E.,
    Stark, A.~A., Bally, J., \& Uchida, K.~I.\ 1991, \mnras, 252, 210 
\bibitem[Bland \& Tully(1989)]{Bland:1989} Bland, J., \& Tully, R.~B.\ 1989,
    \aj, 98, 723
\bibitem[Bland-Hawthorn et al.(2007)]{Bland:2007} Bland-Hawthorn, J.,
    Veilleux, S., \& Cecil, G.\ 2007, \apss, 311, 87
\bibitem[Cooper et al.(2008)]{Cooper:2008} Cooper, J.~L., Bicknell, G.~V.,
    Sutherland, R.~S., \& Bland-Hawthorn, J.\ 2008, \apj, 674, 157
\bibitem[Dopita \& Sutherland(1995)]{Dopita:1995} Dopita, M.~A., \&
    Sutherland, R.~S.\ 1995, \apj, 455, 468
\bibitem[Ferrara \& Tolstoy(2000)]{Ferrara:2000} Ferrara, A., \& Tolstoy, E.\
    2000, \mnras, 313, 291
\bibitem[Forbes et al.(1993)]{Forbes:1993} Forbes, D.~A., Ward, M.~J.,
    Rotaciuc, V., Blietz, M., Genzel, R., Drapatz, S., van der Werf, P.~P.,
    \& Krabbe, A.\ 1993, \apjl, 406, L11
\bibitem[Grevesse et al.(2007)]{Grevesse:2007} Grevesse, N., Asplund, M., \&
    Sauval, A.~J.\ 2007, Space Science Reviews, 130, 105
\bibitem[Hoopes et al.(1996)]{Hoopes:1996} Hoopes, C.~G., Walterbos, R.~A.~M.,
    \& Greenwalt, B.~E.\ 1996, \aj, 112, 1429
\bibitem[Kauffmann et al.(2003)]{Kauffmann:2003} Kauffmann, G., et al.\ 2003,
    \mnras, 346, 1055
\bibitem[Kennicutt(1998)]{Kennicutt:1998} Kennicutt, R.~C., Jr.\ 1998, \araa,
    36, 189
\bibitem[Kewley et al.(2001)]{Kewley:2001} Kewley, L.~J., Dopita, M.~A.,
    Sutherland, R.~S., Heisler, C.~A., \& Trevena, J.\ 2001, \apj, 556, 121
\bibitem[Kobayashi et al.(2007)]{Kobayashi:2007} Kobayashi, M.~A.~R., Totani,
    T., \& Nagashima, M.\ 2007, \apj, 670, 919
\bibitem[Leitherer et al.(1999)]{Leitherer:1999} Leitherer, C., et al.\ 1999,
    \apjs, 123, 3
\bibitem[Liang et al.(2006)]{Liang:2006} Liang, Y.~C., Yin, S.~Y., Hammer, F.,
    Deng, L.~C., Flores, H., \& Zhang, B.\ 2006, \apj, 652, 257
\bibitem[McCarthy et al.(1987)]{McCarthy:1987} McCarthy, P.~J., van Breugel,
    W., \& Heckman, T.\ 1987, \aj, 93, 264
\bibitem[Miller \& Mathews(1972)]{Miller:1972} Miller, J.~S., \& Mathews,
    W.~G.\ 1972, \apj, 172, 593
\bibitem[Moustakas \& Kennicutt(2006)]{Moustakas:2006} Moustakas, J., \&
    Kennicutt, R.~C., Jr.\ 2006, \apj, 651, 155
\bibitem[Osterbrock \& Ferland(2006)]{Osterbrock:2006} Osterbrock, D.~E., \&
    Ferland, G.~J.\ 2006, Astrophysics of gaseous nebulae and active galactic
    nuclei, 2nd.~ed.~by D.E.~Osterbrock and G.J.~Ferland.~Sausalito, CA:
    University Science Books, 2006
\bibitem[Pagel(1997)]{Pagel:1997} Pagel, B.~E.~J.\ 1997, Nucleosynthesis and
    Chemical Evolution of Galaxies, by Bernard E.~J.~Pagel, pp.~392.~ISBN
    0521550610.~Cambridge, UK: Cambridge University Press, October 1997
\bibitem[Paglione et al.(2004)]{Paglione:2004} Paglione, T.~A.~D., Yam, O.,
    Tosaki, T., \& Jackson, J.~M.\ 2004, \apj, 611, 835
\bibitem[Pence(1980)]{Pence:1980} Pence, W.~D.\ 1980, \apj, 239, 54
\bibitem[Peng et al.(1996)]{Peng:1996} Peng, R., Zhou, S., Whiteoak, J.~B., Lo,
    K.~Y., \& Sutton, E.~C.\ 1996, \apj, 470, 821
\bibitem[Rekola et al.(2005)]{Rekola:2005} Rekola, R., Richer, M.~G., McCall,
    M.~L., Valtonen, M.~J., Kotilainen, J.~K., \& Flynn, C.\ 2005, \mnras, 361,
    330
\bibitem[Sakamoto et al.(2006)]{Sakamoto:2006} Sakamoto, K., et al.\ 2006,
    \apj, 636, 685
\bibitem[Salpeter(1955)]{Salpeter:1955} Salpeter, E.~E.\ 1955, \apj, 121, 161
\bibitem[Sams et al.(1994)]{Sams:1994} Sams, B.~J., III, Genzel, R., Eckart,
    A., Tacconi-Garman, L., \& Hofmann, R.\ 1994, \apjl, 430, L33
\bibitem[Sato et al.(2008)]{Sato:2008} Sato, T., Martin, C.~L., Noeske, K.~G.,
    Koo, D.~C., \& Lotz, J.~M.\ 2008, ArXiv e-prints, 804, arXiv:0804.4312
\bibitem[Schulz \& Wegner(1992)]{Schulz:1992} Schulz, H., \& Wegner, G.\ 1992,
    \aap, 266, 167
\bibitem[Shapley et al.(2003)]{Shapley:2003} Shapley, A.~E., Steidel, C.~C.,
    Pettini, M., \& Adelberger, K.~L.\ 2003, \apj, 588, 65
\bibitem[Shopbell \& Bland-Hawthorn(1998)]{Shopbell:1998} Shopbell, P.~L., \&
    Bland-Hawthorn, J.\ 1998, \apj, 493, 129
\bibitem[Strickland et al.(2000)]{Strickland:2000} Strickland, D.~K., Heckman,
    T.~M., Weaver, K.~A., \& Dahlem, M.\ 2000, \aj, 120, 2965
\bibitem[Strickland et al.(2004)]{Strickland:2004} Strickland, D.~K., Heckman,
    T.~M., Colbert, E.~J.~M., Hoopes, C.~G., \& Weaver, K.~A.\ 2004, \apj,
    606, 829
\bibitem[Sugai et al.(2000)]{Sugai:2000} Sugai, H., et al.\ 2000, \procspie,
    4008, 558
\bibitem[Sugai et al.(2002)]{Sugai:2002} Sugai, H., Ozaki, S., Hattori, T., \&
    Kawai, A.\ 2002, Galaxies: the Third Dimension, 282, 433
\bibitem[Sugai et al.(2003)]{Sugai:2003} Sugai, H., Davies, R.~I., \& Ward,
    M.~J.\ 2003, \apjl, 584, L9
\bibitem[Sugai et al.(2004)]{Sugai:2004a} Sugai, H., et al.\ 2004, \procspie,
    5492, 651
\bibitem[Sugai et al.(2004)]{Sugai:2004b} Sugai, H., et al.\ 2004, \apjl, 615,
    L89
\bibitem[Tapken et al.(2007)]{Tapken:2007} Tapken, C., Appenzeller, I., Noll,
    S., Richling, S., Heidt, J., Meink{\"o}hn, E., \& Mehlert, D.\ 2007, \aap,
    467, 63
\bibitem[T{\"u}llmann et al.(2006)]{Tullmann:2006} T{\"u}llmann, R., Pietsch,
    W., Rossa, J., Breitschwerdt, D., \& Dettmar, R.-J.\ 2006, \aap, 448, 43
\bibitem[Turner \& Ho(1985)]{Turner:1985} Turner, J.~L., \& Ho, P.~T.~P.\ 1985,
    \apjl, 299, L77
\bibitem[Ulvestad \& Antonucci(1997)]{Ulvestad:1997} Ulvestad, J.~S., \&
    Antonucci, R.~R.~J.\ 1997, \apj, 488, 621
\bibitem[Veilleux \& Rupke(2002)]{Veilleux:2002} Veilleux, S., \& Rupke, D.~S.\
    2002, \apjl, 565, L63
\bibitem[Veilleux et al.(2005)]{Veilleux:2005} Veilleux, S., Cecil, G., \&
    Bland-Hawthorn, J.\ 2005, \araa, 43, 769
\bibitem[Watson et al.(1996)]{Watson:1996} Watson, A.~M., et al.\ 1996, \aj,
    112, 534
\bibitem[Whitford(1958)]{Whitford:1958} Whitford, A.~E.\ 1958, \aj, 63, 201

\end{thebibliography}
\end{document}